\documentclass[aps,prb,preprint,amsmath,amssymb,showpacs,superscriptaddress]{revtex4-1}
\usepackage{multirow}
\usepackage{graphicx,epsfig}
\usepackage{wasysym}
\usepackage{color,soul} 
\usepackage[colorlinks=true,citecolor=blue,linkcolor=red,linktocpage=true,pagebackref=false]{hyperref}

\def \be{\begin{equation}}
\def \ee {\end{equation}}
\newcommand{\ba}{\begin{eqnarray}}
\newcommand{\ea}{  \end{eqnarray}}
\def \veps {\varepsilon}
\def \calg {{\cal{G}}}

\begin{document}

\title{Time resolved heat exchange in driven quantum systems}

\author{Mar\'{\i}a Florencia Ludovico}
\affiliation{Departamento de F\'isica, FCEyN, Universidad de Buenos Aires and IFIBA, Pabell\'on I, Ciudad Universitaria, 1428 CABA Argentina}
\author{Jong Soo Lim}
\affiliation{School of Physics, Korea Institute for Advanced Study, Seoul 130-722, Korea}
\author{Michael Moskalets}
\affiliation{Department of Metal and Semiconductor Physics,
NTU "Kharkiv Polytechnic Institute", 61002 Kharkiv, Ukraine}
\author{Liliana Arrachea}
\affiliation{Departamento de F\'isica, FCEyN, Universidad de Buenos Aires and IFIBA, Pabell\'on I, Ciudad Universitaria, 1428 CABA Argentina}
\author{David S\'anchez}
\affiliation{Instituto de F\'isica Interdisciplinar y Sistemas Complejos
IFISC (UIB-CSIC), E-07122 Palma de Mallorca, Spain}


\begin{abstract}
We study time-dependent heat transport in systems composed of a resonant level periodically forced with an external power source and coupled to a fermionic continuum.
This simple model contains the basic ingredients to understand time resolved energy exchange in quantum capacitors that behave as single particle emitters.
We analyse the behaviour of the dynamic heat current for driving frequencies within the non-adiabatic regime, showing that it does not obey a Joule dissipation law.
\end{abstract}
\maketitle

\section{Introduction}
 
Recent progress in the miniaturisation of electronic circuits puts on agenda the development of new theoretical methods. 
The energy transport in systems on the nano- and already molecular and atomic scales cannot be treated classically but requires a quantum-mechanical treatment~\cite{Jezouin,but93a,but93b}.
In particular, time-dependent quantum transport in systems that act as quantum capacitors have recently received a lot of attention both experimentally and theoretically \cite{gab06,mos1,mos2,mos3}. Recent experiments show that
a quantum dot tunnel-coupled to a reservoir can be used for on-demand single-electron injection \cite{feb07}.
Electron emission and absorption are periodically generated by applying an external time-periodic potential.
Harmonic potentials are also crucial for the creation of
directed transport in asymmetric systems, such as charge~\cite{lin99} and spin~\cite{cos10} ratchets.
Furthermore, ac fields are shown to control matter tunneling in Bose-Einstein condensates~\cite{lig07}.

Electrons in periodically driven coherent conductors carry energy in addition to current. Thus, the role of electron-like and hole-like excitations created by ac driving in the energy current noise is investigated in Ref. \cite{fra}. Heat production in nanoscale
engines is analyzed in Refs.~\cite{arr07,mos09} while Ref.~\cite{lim13} finds a universal thermal resistance 
for the low-temperature dynamical transport.

In a recent work \cite{dynamical}, we studied the time-resolved energy production and redistribution in ac-driven quantum coherent electron systems. We showed that the coupling between the different parts of the system contributes to the energy transport, and that this contribution is of ac nature. We also presented an appropriate definition for the time resolved heat current in accordance with the fundamental principles of thermodynamics. Interestingly, we showed that for low frequencies of the driving potential (adiabatic regime) we found that the time-dependent heat flux is instantaneously given by the Joule law with a universal resistance. The purpose of this work is to explore if this behaviour of the time resolved heat current remains also valid  for higher driving frequencies beyond the adiabatic regime. To this end, we consider a single dot connected to a fermionic band of continuous density of states (a reservoir) and driven with a harmonically time-dependent potential with a frequency within the non-adiabatic regime.

\section{Model and theoretical treatment}
\begin{figure}[htp]
\begin{center}
\includegraphics[width=0.5\textwidth]{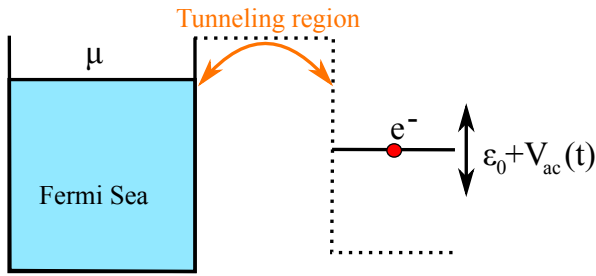}
\caption{Sketch of the setup under consideration. A single electron level is connected to a fermionic band (reservoir with chemical potential $\mu$) via a tunnel barrier. Energy is supplied to the system by a time periodic power source $V_{ac}(t)$ with characteristic frequency $\Omega$. }
\label{setup}
\end{center}
\end{figure}

The system under consideration is sketched in Fig.~\ref{setup}, in which a resonant level is driven harmonically by a power source.  The Hamiltonian of the setup is $H=H_C+H_T+H_D$, where the first term, $H_C=\sum_k\veps_k c_k^\dagger c_k$, represents the fermionic continuum (reservoir) with the energy band $\veps_k$. 
The Hamiltonian of a tunneling region reads $H_T=\sum_k (w_k d^\dagger c_k +h.c)$, where $w_k$ is the coupling amplitude. 
The term  $H_D(t)=\veps_d (t) d^{\dagger} d$ denotes a driven system, in which the energy level, $\veps_d(t)=\veps_0+V_{ac}\,{\sin}(\Omega t)$, is varied in time by the power source. 
For simplicity, we consider noninteracting, spinless electrons.

In order to define the energy fluxes entering each part of the system, we analyze the evolution in time of the total energy,
\be\label{energy}
\frac{d\langle H\rangle}{dt}=J^E_C(t)+J^E_T(t)+J^E_D(t)+P(t),
\ee
 where we identify the  energy flux entering  the reservoir  $J^E_C=i\langle\left[H,H_C\right]\rangle/\hbar$, the change per unit time of the energy stored in the tunneling region $J^E_T=i\langle\left[H,H_T\right]\rangle/\hbar$ and the energy flux entering the resonant level $J^E_D=i\langle\left[H,H_D\right]\rangle/\hbar$. The total Hamiltonian conserves the number of particles but not the total energy due to the power developed by the ac forces $P(t)=d\langle H\rangle/dt=\langle{\partial H_D}/{\partial t}\rangle= n_d(t) d\veps_d(t)/dt $, where $n_d(t)=\langle d^\dagger(t)d(t)\rangle$ is a resonant level occupation probability, i.e., the time-average of the number of electrons present on the resonant level. 
 
 We solve the problem with the non-equilibrium Green function procedure of Refs. \cite{lb,lm}. The different energy fluxes can be computed in terms of the retarded Green function $\calg^r(t,t')=-i\theta(t-t')\langle\{d(t),d^\dagger(t')\}\rangle$, and we use the Floquet-Fourier representation
\be
\calg^r(t,t')=\sum_{n=-\infty}^\infty\int\frac{d\omega}{2\pi}e^{-in\Omega t}e^{-i\omega(t-t')}\calg^r(n,\omega),
\ee
 with $\Omega$ being the fundamental driving frequency. The Floquet components $\calg^r(n,\omega)$ can be computed in term of the equilibrium Green function $\calg^0(\omega)=\left[\omega-\veps_0+i\Gamma/2\right]^{-1}$, where $\Gamma$ represent the hybridization between the electron level and the reservoir. We find that the energy current entering the reservoir at time $t$ reads
\ba\label{cont}
J^E_C(t)  & = & -\sum_l e^{-i l \Omega t} \int \frac{d\veps}{h}  
\{   
i {{\cal G}^r}^{*}(-l,\veps) {\Gamma}
 [(\veps -l \hbar\Omega) f(\veps-l \hbar\Omega)  -
\veps f(\veps)  ] - \nonumber \\
& & \sum_{n} 
[(\veps + \frac{l\hbar\Omega}{2}) f(\veps-n\hbar\Omega)- \veps f(\veps)]
  {\cal G}^r(l+n,\veps-n\hbar\Omega){\Gamma}^2
{{\cal G}^r}^{*}(n,\veps-n\hbar\Omega) \},
\ea
where $f(\veps)=1/[1+e^{(\veps-\mu)/{k_BT}}]$ is the Fermi- Dirac distribution. Following the same procedure, we can compute the other fluxes entering Eq. (\ref{energy}).
The variation of the energy stored in the tunneling region reads
\ba
\label{diff}
J^E_T(t) 
& = & \int\frac{d\veps}{h}\Omega f(\veps)\sum_n n\,2\mbox{Im}\{e^{-i n\Omega t}{\cal{G}}^r(n,\veps){\Gamma}\}.
\ea
Finally, the energy flux entering the dot with a single resonant level can be expressed as  $J^E_D= \veps_d(t)dn_d(t)/dt = -\veps_d(t)J_C(t)/e$. In this expression, we have used charge conservation
to rewrite $dn_d(t)/dt$ in terms of the reservoir current ($J_C(t)= - e dn_d(t)/dt$). 
Applying the above method, the charge current can be expressed as
\ba
\label{charge}
J_C(t)  & = & -\frac{e}{h}\sum_l e^{-i l \Omega t} \int {d\veps}  
\{   
i {{\cal G}^r}^{*}(-l,\veps) {\Gamma}
 [ f(\veps-l\hbar \Omega)  -
f(\veps)  ] - \nonumber \\
& & \sum_{n} 
[ f(\veps-n\hbar\Omega)-  f(\veps)]
  {\cal G}^r(l+n,\veps-n\hbar\Omega){\Gamma}^2
{{\cal G}^r}^{*}(n,\veps-n\hbar\Omega) \}.
\ea
We stress that Eq. (\ref{cont}), (\ref{diff}) and (\ref{charge}) are general and valid up to any order in $\Omega$ and
the ac amplitude $V_{ac}$.

The dc components  of the currents  $\overline{J^E_{C}}$  and $\overline{J^E_{D}}$  satisfy 
$\overline{J^E_{C}}=-\overline{J^E_{D}}$.  The energy conservation expressed by Eq. (\ref{energy})  implies that knowledge on how energy is absorbed or emitted in the contact region, $J^E_{T}(t)$, plays an important role in the definition of heat in the time domain, even though this term  satisfies $\overline{J^E_{T}}=0$. The dc component of heat flowing into the reservoir is related to energy and charge currents as follows, $\overline{J^Q_{C}}=\overline{J^E_{C}}-\mu \overline{J_{C}}$. In Ref. \cite{dynamical}
we focused on the  adiabatic regime and adopted a thermodynamical analysis to define heat, based on the fact  that the reservoir is a macroscopic system, which experiences little changes  under slow variations due to the driving at the quantum dot.  Thus, the rate of change of the internal energy in the reservoir leads to the appropriate definition of heat exchange between the reservoir and the driven part of the system,
\be
\label{heat}
J^Q(t)=J^E_C(t)+J^E_T(t)/2-\mu J_C(t),
\ee  
where we can interpret the quantity $J^E_T(t)$ as the chemical work developed by the contact when an electron is flowing through it. 

The above definition of the heat flow in the time domain within the adiabatic regime is fully consistent with the treatment based on the scattering matrix formalism, as shown in \cite{dynamical}. Furthermore,
it was also shown that such definition leads to a generalized Joule law in the time domain within the adiabatic regime
\be \label{joule}
J^Q(t)=R_q [J_C(t)]^2  ,
\ee
where $R_q=h/2e^2$ is the universal resistance of the contact. 
This law reflects the fact that within the adiabatic regime the heat generated by an ac driving flows into the reservoir and increases its entropy at {\em every time}.

\section{Results}
\begin{figure}[htp]
\begin{center}
\includegraphics[width=0.8\textwidth]{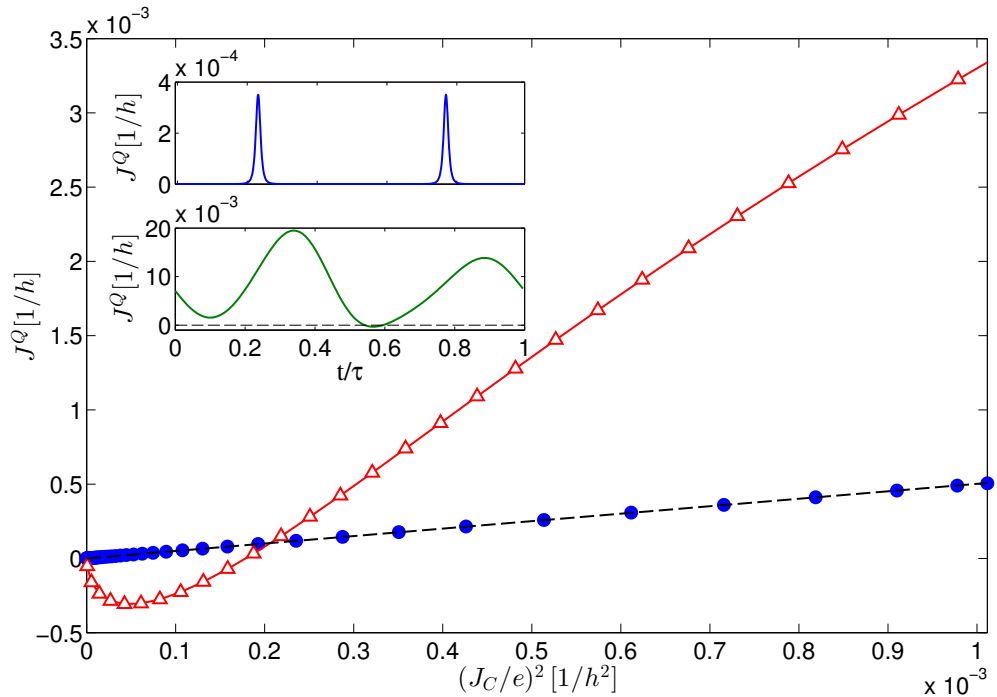}
\caption{Heat fluxes $J^{Q}(t)$ as a function of the charge current $J_C^{2}(t)$ for two different driving frequencies $\Omega$. Open triangles corresponds to a frequency within the non-adiabatic regime $\hbar\Omega=0.3$, $\mu=0.2$, $\veps=0$, $V_{ac}=0.6$, and $T=0$. Solid circles corresponds to a slow driving frequency $\hbar\Omega=10^{-3}$, $\mu=0$, $\veps=-1.2$, $V_{ac}=10$, and $T=0$. Energies are expressed in units of $\Gamma$. Dashed line represents the Joule dissipation relation with a resistance $R=h/2e^2$.  
Inset: Upper panel shows $J^{Q}(t)$ within the slow driving regime; lower panel depicts $J^{Q}(t)$ for the non-adiabatic regime. }
\label{fig:figura1}
\end{center}
\end{figure}

The question arises now to what extent the definition of both a time-resolved heat current (\ref{heat}) and a corresponding time-dependent Joule law (\ref{joule}) remain valid when the driving frequency is high and the transport is non-adiabatic.  
In order to explore this issue we have explicitly evaluated time-dependent heat and charge currents $J^Q(t)$ and $J_C(t)$ for different driving frequencies $\Omega$. Results  are shown in Fig. \ref{fig:figura1}, where, for simplicity, we have considered zero temperature ($T=0$). We present results for slow driving frequencies (solid circles), where the heat current is linear as a function of $J_C^2(t)$, with the universal slope $R_q$. 
On the other hand, we show (open triangles) that the induced heat current for higher frequencies (non-adiabatic regime) departs from the Joule law. Furthermore it may attain negatives values for some times,  which is in seeming contradiction with the second law of the thermodynamics. This puzzling result, a  consequence of system's non-equilibrium dynamics, raises a question about the correct formulation of the second law of the thermodynamics for strongly nonequilibrium systems.
At slow driving frequencies the heat current behaves as expected for stationary systems (remains positive and obeys a Joule dissipation law), for which the classical laws of thermodynamics remain valid. However, at high driving frequencies, the non-equilibrium and quantum effects may become relevant.  In particular, the uncertainty relation between time and energy may play a role. This feature deserves a more detailed analysis. 

\section{Summary and conclusions}

In conclusion, we have discussed dynamical heat generation in a resonant level system due to coupling to an external time-dependent potential with a typical frequency within the non-adiabatic regime. Unlike a slow driving frequency regime, where the time-dependent heat flux is instantaneously given by the Joule law with universal resistance, for higher driving frequencies the heat flux can attain negative values, which may arise from a non-equilibrium dynamics. Further analysis is necessary to better understand this phenomenon.

\section{Acknowledgment}
We acknowledge support from CONICET, MINCYT and UBACYT, Argentina (MFL and LA), MINECO Grant No.~FIS2011-23526,
CAIB and FEDER.


\section*{References}
\begin{thebibliography}{99}

\bibitem{Jezouin}
S. Jezouin, F. D. Parmentier, A. Anthore, U. Gennser, A. Cavanna,
Y. Jin, and F. Pierre, Science {\bf 343}, 601 (2013).
\bibitem{but93a}
M. B\"uttiker, A. Pr\^etre, and H. Thomas,
Phys. Lett. A {\bf 180}, 364 (1993).

\bibitem{but93b}
M. B\"uttiker, A. Pr\^etre, and H. Thomas,
Phys. Rev. Lett. \textbf{70}, 4114 (1993). 
\bibitem{gab06}
J.~Gabelli, G.~F\`eve, J.-M.~Berroir, B.~Pla\c cais, A.~Cavanna, B.~Etienne,
Y.~Jin, and D.C.~Glattli, Science {\bf 313}, 49 (2006).

\bibitem{mos1} J. Splettstoesser, S. Ol'khovskaya, M. Moskalets and M. B\"uttiker, Phys. Rev. B \textbf{78}, 205110 (2008)
\bibitem{mos2} M. Moskalets and M. B\"uttiker, Phys. Rev. B \textbf{80}, 081302(R) (2009)
\bibitem{mos3} M. Moskalets, P. Samuelsson and M. B\"uttiker, Phys. Rev. Lett. \textbf{100}, 086601 (2008).
\bibitem{feb07}
G. F\`eve, A. Mah\'e A, J.M. Berroirm, T. Kontos, B.~Pla\c cais, D.C.~Glattli,
A.~Cavanna, B.~Etienne and Y.~Jin, Science \textbf{316} 1169 (2007).
\bibitem{lin99}
H. Linke, T.E. Humphrey, A. L\"ofgren, A. Sushkov, R. Newbury, R.P. Taylor, and P.Omling,
Science \textbf{286}, 2314 (1999).
\bibitem{cos10}
M.V. Costache and S.O. Valenzuela,
Science \textbf{330}, 1645 (2010).
\bibitem{lig07}
H. Lignier, C. Sias, D. Ciampini, Y. Singh, A. Zenesini, O. Morsch, and E. Arimondo,
Phys. Rev. Lett. \textbf{99}, 220403 (2007).
\bibitem{fra} F. Battista, F. Haupt and J. Splettstoesser, arXiv:1405.4326 (preprint).
\bibitem{arr07}
L. Arrachea, M. Moskalets, and L. Martin-Moreno,
Phys. Rev. B \textbf{75}, 245420 (2007).
\bibitem{mos09}
M. Moskalets and M. B\"uttiker,
Phys. Rev. B \textbf{80}, 081302(R) (2009).
\bibitem{lim13}
J.S. Lim, R. L\'opez and D. S\'anchez,
Phys. Rev. B  \textbf{88}, 201304(R) (2013).
\bibitem{dynamical}
M.F. Ludovico, J.S. Lim, M. Moskalets, L. Arrachea, and D. S\'anchez, Phys. Rev B \textbf{89}, 161306(R) (2014).

\bibitem{lb}
 L. Arrachea, Phys. Rev. B \textbf{66} (2002) 045315; L. Arrachea, Phys. Rev. B \textbf{70} (2004) 155407
 
 \bibitem{lm}
L. Arrachea, Phys. Rev. B \textbf{72} (2005) 125349;
L. Arrachea, M. Moskalets, Phys. Rev. B \textbf{74} (2006) 245322;
L. Arrachea, Phys. Rev. B \textbf{75} (2007) 035319.

\bibitem{balian}
R. Balian, {\em From Microphysics to Macrophysics}, Vol. I,
Springer, Berlin-Heidelberg, 1991-2007.


\end{thebibliography}
\end{document}